\documentclass[11pt]{article}
\usepackage{times,aaspp4,epsf,flushrt}

\input{epsf}

\let\oldfootsep=\footnotesep
\def\lsim{\hbox{ \rlap{\raise 0.425ex\hbox{$<$}}\lower 0.65ex\hbox{$\sim$} }}
\def\gsim{\hbox{ \rlap{\raise 0.425ex\hbox{$>$}}\lower 0.65ex\hbox{$\sim$} }}

\def\etal{et~al.}

\def\msun { \rm {M_\odot}}
\def\rsun { \rm {R_\odot}}

\def\umin{u_{\rm min}}

\def\tstar{t_{\rm *}}

\def\umin{u_{\rm min}}
\def\t0{t_{\rm 0}}

\def\pac{Paczy{\'n}ski }
\def\ie{{\it i.e. }}

\def\kms {\,{\rm km \, s^{-1} }}
\def\kpc {\, {\rm kpc}}

\def\spose#1{\hbox to 0pt{#1\hss}}
\def\simlt{\mathrel{\spose{\lower 3pt\hbox{$\mathchar"218$}}
     \raise 2.0pt\hbox{$\mathchar"13C$}}}
\def\simgt{\mathrel{\spose{\lower 3pt\hbox{$\mathchar"218$}}
     \raise 2.0pt\hbox{$\mathchar"13E$}}}

\newcommand{\cit}[1]{$^{\ref{#1}}$}
\newcommand{\citt}[2]{$^{\ref{#1},\ref{#2}}$}
\newcommand{\cittt}[3]{$^{\ref{#1},\ref{#2},\ref{#3}}$}
\newcommand{\citttt}[4]{$^{\ref{#1},\ref{#2},\ref{#3},\ref{#4}}$}

\begin{document}

\begin{center}

\title{
            Gravitational Microlensing Evidence for
            a Planet Orbiting a Binary Star System
}
\end{center}
\medskip

\begin{center}
D.P.~Bennett\rlap,$^{\ast}$ S.H.~Rhie\rlap,$^{\ast}$ 
A.C.~Becker\rlap,$^{\dagger}$ N.~Butler\rlap,$^{\ast}$ J.~Dann\rlap,$^\ddagger$ 
S.~Kaspi\rlap,$^\ddagger$ E.M.~Leibowitz\rlap,$^\ddagger$
Y.~Lipkin\rlap,$^\ddagger$ D.~Maoz\rlap,$^\ddagger$
H.~Mendelson\rlap,$^\ddagger$ B.A.~Peterson\rlap,$^\parallel$
J.~Quinn\rlap,$^{\ast}$ O.~Shemmer\rlap,$^{\ddagger}$ S.~Thomson\rlap,$^{\#}$ 
S.E.~Turner$^{\ast\ast}$
\end{center}
\begin{center}
(The Microlensing Planet Search Collaboration and 
 the Wise Observatory GMAN Team)
\end{center}

\begin{tabular}{ll}
$\ast$ & Physics Department, University of Notre Dame, Notre Dame, IN 46530,
         USA \\
$\dagger$ & Departments of Physics and Astronomy, 
    Univ.~of Washington, Seattle, WA 98195, USA \\
$\ddagger$ & School of Physics \& Astronomy and Wise Obs.,
           Tel-Aviv Univ., Tel-Aviv 69978, Israel \\
$\parallel$ & Mt.  Stromlo and Siding Spring Obs.,
           Australian National Univ., Weston, ACT 2611, Australia \\
$\#$ & Department of Mathematics \& Statistics, 
           Monash Univ., Clayton, Victoria 3168, Australia \\
$\ast\ast$ & Department of Physics, University of California, 
Los Angeles, CA, USA \\
\end{tabular}
\bigskip

\parskip 0pt  

{\bf
The study of extra-solar planetary systems has emerged as a new discipline
of observational astronomy in the past few years with the discovery of a 
number of extra-solar planets\rlap.\citt{mq-51peg}{marcy-but-rev}
The properties of most of these extra-solar planets were not anticipated
by theoretical work on the formation of planetary systems,
although the radial velocity technique used for these discoveries is
not yet sensitive to planetary systems like our own\rlap.\cit{cumming_etal}
Here we report observations and light curve modeling of gravitational
microlensing event MACHO-97-BLG-41, which indicates that the lens system
consists of a planet orbiting a binary star system.
According to this model, the mass ratio of 
the binary star system is 3.8:1 and the stars are most likely
to be a late K dwarf and an M dwarf with a separation of about 1.8 AU.
A planet of about 3 Jupiter masses orbits this system at a distance of about
7 AU. If our interpretation of this light curve is correct, it
represents the discovery of a planet orbiting a binary star system
and the first detection of a Jovian planet via the gravitational microlensing
technique. It suggests that giant planets may be 
common in short period binary star systems.
}

The Microlensing Planet Search (MPS) Collaboration aims to detect planets
that orbit distant stars by detecting the influence of 
planets on gravitational microlensing event light curves. Gravitational
microlensing events are observed via the time varying magnification of
the sum of the gravitationally lensed images as the lens system passes
in front of the background source star. The separation of the images is too
small to observe with current instruments. A planet orbiting the lens star
can be detected via a brief deviations of the microlensing light 
curve\cit{mao-pac} from the normal single lens light 
curve\rlap.\cittt{eros-1st}{macho-gpe}{ogle-1st}

The microlensing event MACHO-97-BLG-41 was discovered by the MACHO
team\cit{macho-binaries} and announced on 19 June, 1997. 
MPS observations from the Mt.~Stromlo
1.9m telescope began on that night.  On 29 June, MACHO issued a further 
announcement that the light curve of this event did not have the shape 
expected for a single lens event, and the PLANET team issued a similar
announcement on 2 July.
Regular observations by the GMAN follow-up team began shortly after this
second MACHO announcement with nightly observations from the CTIO 0.9m
telescope and less frequent observations from the Wise Observatory 1.0m
telescope. The MACHO and GMAN data were previously presented without
analysis by the MACHO/GMAN group\rlap.\cit{macho-binaries}. Here,
we present a combined analysis of the MPS and MACHO/GMAN data.

The modeling of MACHO-97-BLG-41 has proven to be more difficult than
any other multiple lensing event observed by MACHO/GMAN or MPS to date.
Figure \ref{fig-lc} shows the light curve of the combined data set along
with the best fit model, while Figure \ref{fig-caustic} shows the trajectory
of the source star with respect to the lens masses and the overall
lens system caustic structure. Most of the features of these multiple
lens light curves can be understood by inspection of diagrams like
Figure \ref{fig-caustic} because the regions of high magnification
are associated with the caustic curves. The spikes and U-shaped
light curves that commonly occur in binary lens lightcurves\cit{macho-binaries}
are easily understood as caustic crossing features.

Our initial attempts to fit the MACHO-97-BLG-41 light curve focused on
both static and orbiting binary lens models. Static binary lens 
models have been successful in fitting the light curves of all the other 
known multiple lens 
events\rlap,\citttt{macho-binaries}{planet-97blg28}{ogle7}{duo2}
but such a model is unable to account for the two caustic crossing
features of MACHO-97-BLG-41. A static binary lens model would necessarily
have a light curve peak around July 21 that is much larger than what is 
actually observed, because a cusp of the central caustic must point in the
direction of the caustic that was crossed on June 20-21. The inclusion of
binary lens possible orbital motion\cit{dominik} also failed to generate
an acceptable fit.

When we concluded that binary lens models would not fit the data, we turned to
triple lens models (Rhie \& Bennett, in preparation), and our fitting
code quickly converged to a model which appears to correspond to
a stable dynamical system.  The light curve for this model
is shown in Figure \ref{fig-lc}. The model parameters are 
expressed in terms of the Einstein ring radius, $R_E$, which
is the size of the ring image that would be seen with perfect lens-source
alignment (\ie with the source and all the lens masses in a line). For a 
typical Galactic bulge lens system, $R_E\sim 3\,$AU.
The fit parameters are as follows:
The Einstein ring radius crossing time is $t_E = 27.832\,$days. The 
closest approach between the angular positions of the source and the
lens system center of mass is $\umin = 0.0679$ (in units of $R_E$) which
occurs at time $t_0 = 23.593$ July 1999, UT.  The mass fractions of
the three masses are 0.7870, 0.2086, and 0.0044, and the
separations are $0.4822\,R_E$ and $1.768\,R_E$  between the mass pairs
1-2 and 1-3, respectively.  The opening angle between the
mass 1-2 and mass 1-3 vectors is 108.30$^\circ$, and the source trajectory
intersects the lens 1-2 axis at an angle of -64.81$^\circ$. The measured
widths and magnifications at the caustic
crossings allow us to determine the time for the source star to move by its 
own radius with respect to the lens axis: $\tstar = 0.16\,$days,
assuming linear limb darkening parameters of 0.656 for the standard R-band, 
0.632 for the 
MACHO-R band, and 0.777 for the MACHO-V band\cit{limb1} based upon
a spectrum of the 97-BLG-41 source star\rlap.\cit{lennon-esomes} The relative
locations of the lens masses, the caustics, and the source trajectory
are shown in Figure \ref{fig-caustic}. This model also indicates that
10-20\% of the flux (depending on the pass band)
identified with the source star is unlensed, indicating that
one or more unlensed stars lie within 1-2" of the lensed
source. Such blending is quite common in such crowded stellar fields.
Our model has a $\chi^2 = 1102.96$ for 779 degrees of freedom, which
is typical of other microlensing events without binary or planetary 
light curve deviations. The best fit orbiting binary source model has 
a $\chi^2$ that is larger by 67.57 {\it and} an implied orbital 
velocity larger than the escape velocity of the binary system.

Orbital stability is an important concern for triple star systems because
many such systems are dynamically unstable.  Our gravitational microlensing
model indicates that the planet is separated from the stellar system
center of mass by 3.9 times the binary star separation, in the plane
perpendicular to the line of sight. A recent stability study by
Holman and Wiegert\cit{holman-wiegert} indicates that the stability
of planets orbiting binary star systems depends somewhat on the stellar
system orbital eccentricity. For our best fit mass ratio and
assuming a circular stellar orbit, they find that planetary orbits
are stable if their median separation is at least
2.2 times that of the stars while
for an eccentricity as high as 0.7, the critical ratio of the median 
separations climbs to about 3.1. Thus, our triple lens fit appears to
correspond to a stable system as long as the separation ratio is not
substantially reduced when the unobserved line-of-sight component is added.

Further details regarding the properties of the lens system can be
obtained by comparing the source star radius crossing time, $\tstar$, to the
properties of this star as determined by Lennon \etal\cit{lennon-esomes}
Their spectrum indicates
$T_{\rm eff} = 5000\pm 200\,$K, $\log g = 3.2\pm 0.3$, and
$\left[{\rm Fe/H}\right] = -0.2 \pm 0.3\,$dex,  while our model and the
MACHO Project's photometric calibration\cit{macho-calib} indicate
that the unlensed
magnitude and color of the source star are $V = 19.66 \pm 0.20$ and
$V-R = 1.24 \pm 0.10$. This allows us to estimate the extinction to be
$A_V = 2.9\pm 0.4$, and this leads to an
estimate of the angular size of the source star:
$\theta_\ast = 2.9\pm 0.7\mu $arc sec. The distance to the source star is
not known precisely, but it is likely to be at or slightly beyond the
center of the Galaxy at $D \sim 8.5\kpc$. At this distance, the source
star would have a radius of $5.3\pm 1.3\rsun$.  A comparison to the
Bertelli \etal\cit{bertelli} isochrones indicates that a number of possible 
early K class III-IV stellar models are consistent with these parameters. 
These range from a 10 billion year old $1\msun$ star to a 1.5
billion year old star of $1.7\msun$.

Our estimate of the angular size of the source star, combined with
our measurement of the stellar radius crossing time, yields the
relative proper motion of the lens with respect to the source:
$\mu = 6.7\pm 1.7\,$mas/yr. Projected to the likely source distance of
$8.5\kpc$, this is $270\kms$ which is a reasonable value for the relative
transverse velocity between a pair of Galactic bulge stars. 
The measurements of $\mu$ and $t_E$ allow us to
solve for the single parameter family of solutions relating lens mass and
distance shown in Figure \ref{fig-masslike}. This Figure also shows
likelihood function for the lens system location which is based upon
a simple model of Galactic kinematics and our upper limit on the lens
brightness.

The likelihood function in Figure \ref{fig-masslike} implies 
$D_{\rm lens} = 6.3{+0.6\atop -1.3}\,\kpc$ and 
$M_{\rm tot} = 0.8\pm 0.4\msun$. The individual lens masses are
$M_1 = 0.6\pm 0.3\msun$, $M_2 = 0.16\pm 0.08\msun$, and 
$M_3 = 0.0033\pm 0.0017\msun$ (or $M_3 = 3.5 \pm 1.8$ Jupiter masses). 
The implied transverse separations are
$1.5{+0.1\atop -0.3}\,$AU for the two stars, and
$5.7{+0.6\atop -1.1}\,$AU for the planet from the
center of mass of the system. Assuming a random orientation,
the most likely three dimensional separations
are a stellar separation of $1.8\,$AU, and a planetary separation of $7.0\,$AU.
Thus, the most likely lens system consists of a
K-dwarf--M-dwarf binary star pair orbited by a planet of about
3 Jupiter masses.

One alternative explanation of this event might be
a binary source star system lensed by
a binary lens system\rlap.\cit{gaudi} This would require that the
secondary source star be much fainter than the primary, and that it
be lensed by a much greater amount so that the width of the secondary
source light curve will be much narrower than the primary source 
light curve. In general, one would expect a significant color shift
between the secondary and primary peaks in such a situation.
The MACHO data show no evidence for such a color shift.
The PLANET Collaboration apparently has additional data on this
event\cit{glanz} in the I-band and perhaps in the V-band as well, so
future attempts to model the light curve may be able to place tighter
constraints on any color change.  A definitive test of the 
binary source-binary lens hypothesis would be to search for radial 
velocity variability of the source star due to its orbital motion if 
it is indeed a binary.

Our apparent detection of a planet orbiting a binary star system
represents both the first discovery of a Jovian planet via gravitational
microlensing and the first discovery of a planet orbiting a binary
star system. (MPS and MOA\cit{mps-98blg35} have recently reported a 
possible detection of a low mass planet via microlensing). A number of planets
have been discovered orbiting individual members of widely separated
binary systems\rlap,\cit{marcy-but-rev} but these are systems that probably
became binaries by gravitational capture after the planets had formed.
The planet orbiting the
MACHO-97-BLG-41 lens system is likely to be the first example of a planet
that {\it formed} in a binary system although circumbinary disks which
might be in the process of forming planets have been
observed\rlap.\cit{jenmat}. It is somewhat
curious that the first detection of a planet via gravitational microlensing
should apparently be orbiting a binary star system. Such planets have
not been seen in planetary searches using the radial velocity technique
because the radial velocity search programs have avoided short period
binary stars. Microlensing planet search programs tend to concentrate
on binary lensing events because they provide better opportunities for
non-planetary science\rlap.\citt{planet-97blg28}{macho-binaries} Since
these events comprise $\simlt 10\,$\% of the total number of microlensing
events, many more single lens events are observed, and it was
expected that microlensing would detect Jovian planets 
orbiting single stars first. Our present result suggests that 
Jovian planets may be more common in short period binary systems than in
single star systems.



\centerline{\bf REFERENCES}

\begin{enumerate}

\item \label{mq-51peg}
  Mayor, M. \& Queloz, D.
  A Jupiter-Mass Companion to a Solar-Type Star.
  {\sl Nature} {\bf 378}, 355-357 (1995).

\item \label{marcy-but-rev}
  Marcy, G.~W. \& Butler, R.~P.
  Detection of Extrasolar Giant Planets.
  {\sl Ann. Rev. Astronomy and Astrophys.} {\bf 36}, 57 (1998).

\item \label{cumming_etal}
  Cumming, A., Marcy, G.~W. \& Butler, R.~P.
  The Lick Planet Search: Detectability and Mass Thresholds
  {\sl Astrophys. J.}, in press, (1999) astro-ph/9906466.

\item \label{macho-binaries}
  Alcock, C. \etal\
  Binary Microlensing Events from the MACHO Project.
  submitted to {\sl Astrophys. J.}, (1999) astro-ph/9907369.

\item \label{mao-pac}
  Mao, S. \& \pac, B.
  Gravitational Microlensing by Double Stars and Planetary Systems.
  {\sl Astrophys. J.} {\bf 374}, L37-41 (1991).







\item \label{eros-1st}
  Aubourg, E. \etal\
  Evidence for Gravitational Microlensing by Dark Objects in the Galactic Halo
  {\sl Nature} {\bf 365}, 623-625 (1993).

\item \label{macho-gpe}
  Alcock, C. \etal\
  Possible Gravitational Microlensing of a Star in the Large Magellanic Cloud
  {\sl Nature} {\bf 365}, 621-623 (1993).

\item \label{ogle-1st}
  Udalski, A., Szymanski, M., Kaluzny, J., Kubiak, M., Krzeminski, W.
  Mateo, M., Preston, G.~W. \& \pac, B.
  The Optical Gravitational Lensing Experiment. Discovery of the First
  Candidate Microlensing Event in the Direction of the Galactic Bulge.
  {\sl Acta Astronomica} {\bf 43}, 289-293 (1993).





\item \label{sod}
  Bennett, D.~P. \etal\
  The MACHO Project II: Data Reduction and Analysis of 6 Million Lightcurves.
  American Astronomical Society Meeting, {bf 183}, 7206 (1993).

\item \label{allframe}
  Stetson, P. 
  The Center of the Core-cusp Globular Cluster M15: CFHT and HST Observations,
  ALLFRAME Reductions.
  {\sl Proc. Astron. Soc. Pac.}, {\bf 106}, 250-280 (1994).

\item \label{planet-97blg28}
  Albrow, M. \etal\
  Limb-Darkening of a K Giant in the Galactic Bulge: PLANET Photometry
  of MACHO 97-BLG-28.
  {\sl Astrophys. J.} in press (1999), astro-ph/9811479.

\item \label{ogle7}
  Udalski, A. \etal\
  The Optical Gravitational Lensing Experiment: OGLE no. 7: Binary 
  Microlens or a New Unusual Variable?
  {\sl Astrophys. J.} {\bf 436}, L103-106 (1994).

\item \label{duo2}
  Alard, C., Mao, S. \& Guibert, J.
  Object DUO 2: a New Binary Lens Candidate?
  {\sl Astron. Astrophys.} {\bf 300}, L17-20 (1995).

\item \label{dominik}
  Dominik, M.
  Galactic Microlensing with Rotating Binaries.
  {\sl Astron. Astrophys.} {\bf 329}, 361-374 (1998).

\item \label{lennon-esomes}
  Lennon, D.~J., Mao, S., Reetz, J., Gehren, T., Yan, L. \& Renzini, A.
  Real-time Spectroscopy of Gravitational Microlensing Events - Probing the 
  Evolution of the Galactic Bulge.
  {\sl The ESO Messenger} {\bf 90}, 30-38 (1997). (astro-ph/9711147).

\item \label{limb1}
  Diaz-Cordoves, J., Claret, A. \& Gimenez, A.
  Linear and Non-linear Limb-darkening Coefficients for LTE Model Atmospheres.
  {\sl Astron. Astrophys. Suppl. Ser.} {\bf 110}, 329-350 (1995).


\item \label{holman-wiegert}
  Holman, M.~J. and Wiegert, P.~A.
  Long-Term Stability of Planets in Binary Systems.
  {\sl Astron. J.} {\bf 117}, 621-628 (1999).

\item \label{bertelli}
  Bertelli, G., Bressan, A., Chiosi, C., Fagotto, F. \& Nasi, E.
  Theoretical Isochrones from Models with New Radiative Opacities.
  {\sl Astron. Astrophys. Suppl. Ser.} {\bf 106}, 275-302 (1994).

\item \label{macho-calib}
  Alcock, C. \etal\
  Calibration of the MACHO Photometry Database.
  {\sl Proc. Astron. Soc. Pac.} in press, (1999).

\item \label{han-gould}
  Han, C., \& Gould, A.,
  The Mass Spectrum of MACHOs from Parallax Measurements.
  {\sl Astrophys. J.} {\bf 447}, 53-61 (1995).

\item \label{gaudi}
  Gaudi, B.~S.,
  Distinguishing Between Binary-Source and Planetary Microlensing Perturbations.
  {\sl Astrophys. J.} {\bf 506}, 533-539 (1998).

\item \label{mps-98blg35}
  Rhie, S.~H. \etal
  On Planetary Companions to the MACHO-98-BLG-35 Microlens Star.
  {\sl Astrophys. J.}, in press, (1999) astro-ph/9905151.

\item \label{jenmat}
  Jensen, E.~L.~N., \& Mathieu, R.~D., 
  Evidence for Cleared Regions in the Disks Around Pre-Main-Sequence 
  Spectroscopic Binaries {\sl Astron. J.}, 114, 301--316 (1997).

\item \label{glanz}
  Glanz, J.,
  Flickers From Far-Off Planets
  {\sl Science} {\bf 277}, 765, (1997).






\end{enumerate}

\noindent
{\bf Acknowledgments} 
We would like to thank the MACHO, OGLE, and EROS Collaborations for their
announcements of microlensing events in progress, and we are particularly
grateful to MACHO for their discovery of MACHO-97-BLG-41.  We'd also
like to thank Will Sutherland for access to his likelihood estimation code.

This research has been supported in part by the NASA Origins
program, the National Science Foundation, and
by a Research Innovation Award from the Research Corporation.  Work
performed at MSSSO is supported by the Australian Department of Industry,
Technology and Regional Development.  Work performed at U.W.
is supported in part by the Office of Science and Technology
Centers of NSF.  Astronomy at Wise
Observatory is supported by grants from the Israel Science Foundation.
\pagebreak

\figcaption[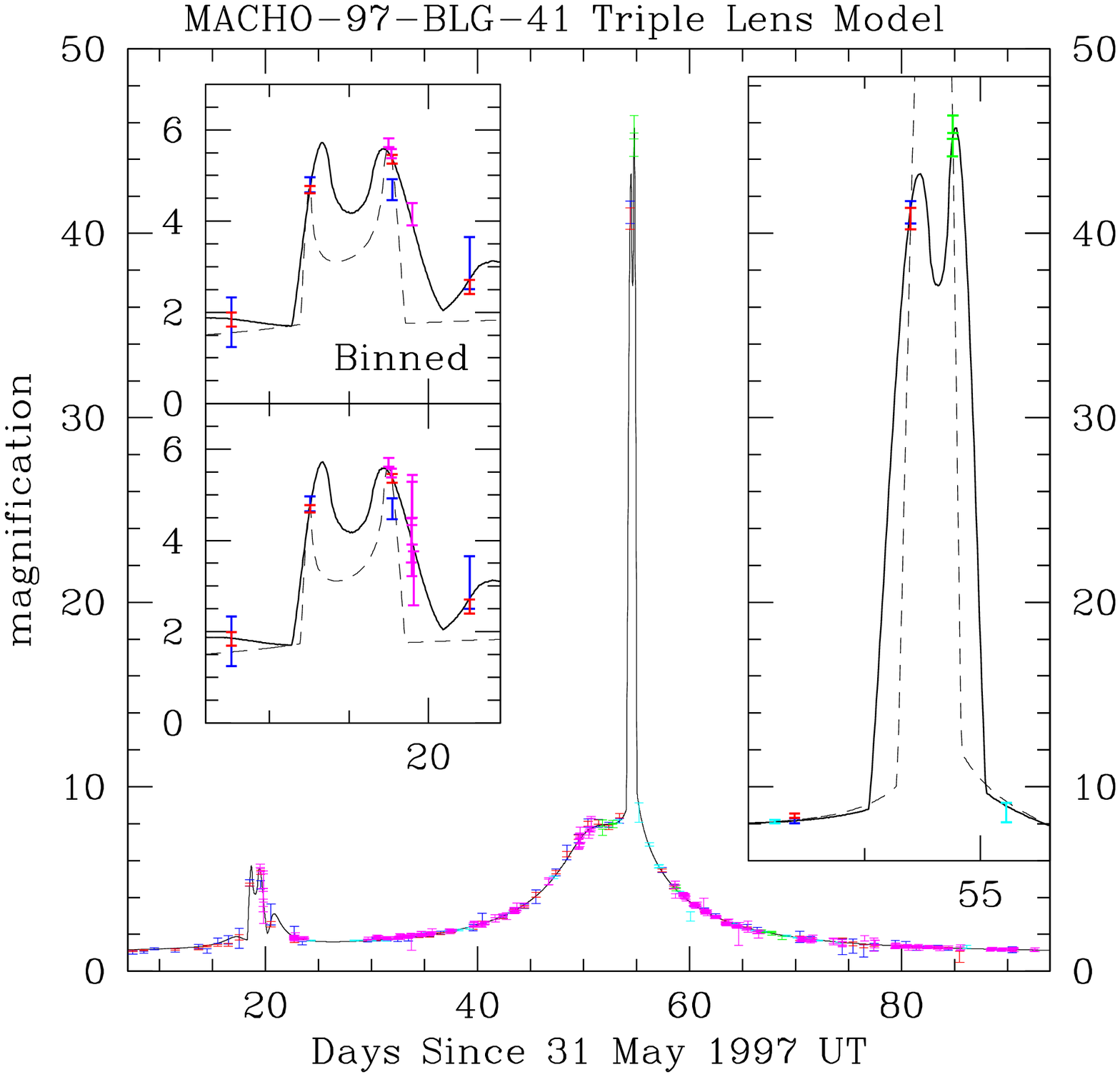]
{The light curve and best fit model for the MACHO-97-BLG-41 
microlensing event are plotted as a function of time. 
The data consist of 356 MPS R-band observations
from the Mt. Stromlo 1.9m telescope, 197 MACHO-R and 194 MACHO-V band
observations from the Mt. Stromlo 1.3m telescope, 35 R-band observations
from the CTIO 0.9m telescope, and 17 R-band observations from the Wise 1.0m
telescope. The MACHO-R, MACHO-V, Wise-R, CTIO-R,
and MPS data are plotted in red, blue, green, cyan, and magenta, respectively.
The inset figures show closeups of the caustic crossing regions of the light
curves, and the tick interval for these figures is 1 day. The last 5 
observations from the night of June 20 were short exposures taken
in bright moonlight over a span of 40 minutes. These have been averaged into
a single data point for the inset figure in the upper left.
The MPS and MACHO data were reduced with the SoDophot photometry
routine\cit{sod} while the CTIO and Wise data were reduced with
ALLFRAME\cit{allframe}.  The MACHO, CTIO, and Wise data were 
previously presented by Alcock \etal\cit{macho-binaries}. The solid curve
is the best fit triple lens model described in the text, and the dashed
curve in the inset figures is the ``best fit" orbiting binary lens light curve
which requires an ``orbital velocity" larger than the 
escape velocity of the binary system and is therefore unphysical.
This large orbital velocity is due to the large distance that the binary
stars must move to account for both caustic crossing features and the
small $\tstar$ value for this fit which is required to achieve the 
magnifications observed on June 20-21.
  \label{fig-lc} }

\figcaption[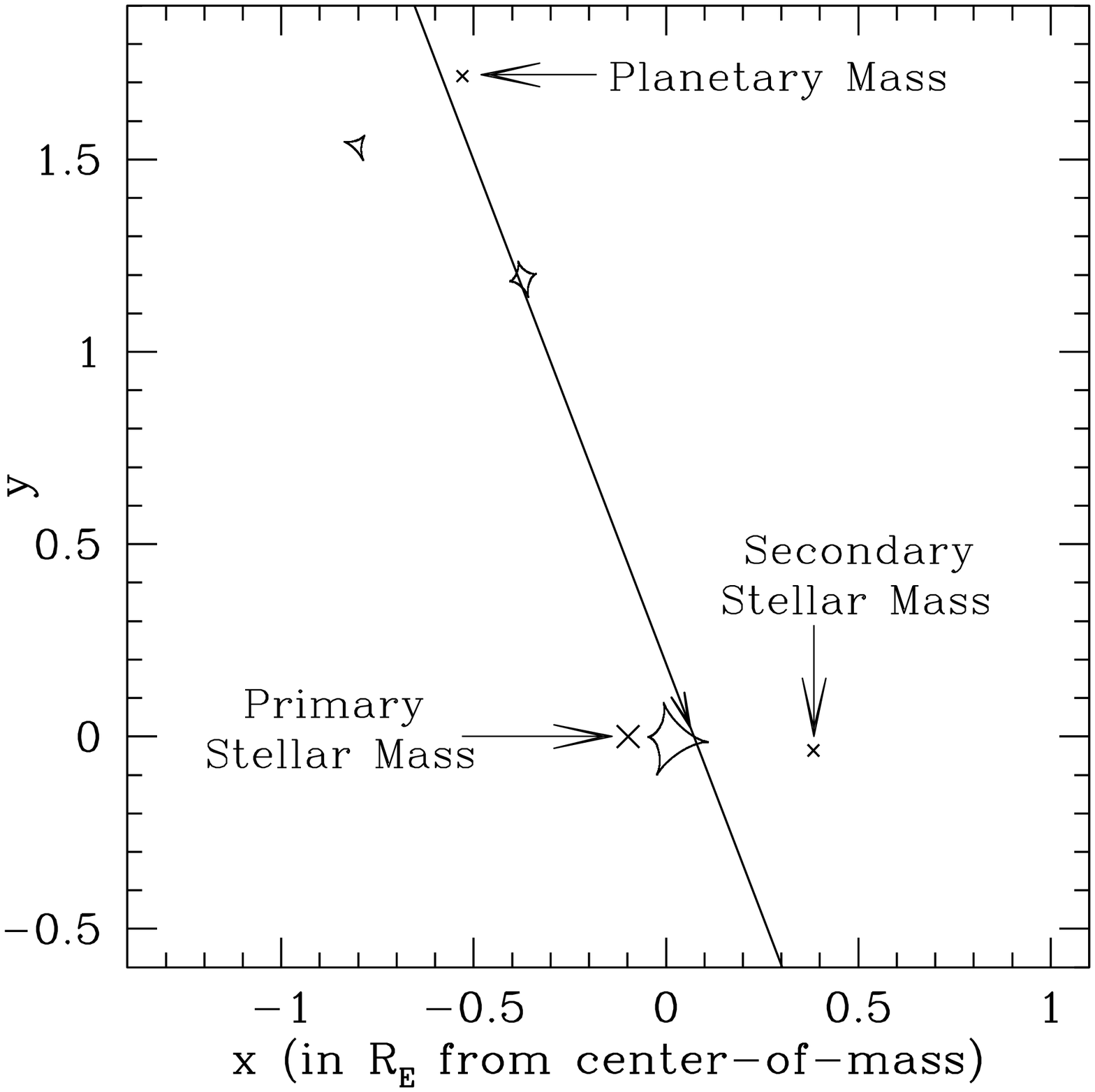]
{The caustic curves for the best fit model for the MACHO-97-BLG-41 
microlensing event are shown along with the lens positions and
the source trajectory which is the diagonal line running from the top
left to the bottom right. High magnification occurs
when the source passes inside a caustic curve where the magnification
is proportional to the inverse square root of the distance to the caustic from
the interior.  High magnification also occurs
outside a caustic curve in the vicinity of a caustic curve cusp.
The observed light curve of MACHO-97-BLG-41
has the following features that can be seen in this
Figure and in Figure \ref{fig-lc}: 1) the passage of an isolated
caustic curve on June 19-20, 2) an approach to a cusp on July 21
that gives rise to the light curve shoulder, 3) an extremely
bright spike indicating the crossing of a very narrow caustic
structure during July 24-25. The triangular caustic curve located near the
planetary mass is due to the two stellar mass lenses and has no effect on
the light curve although a similar caustic does play a role in the 
unphysical orbiting binary lens fit. The orbiting binary lens fit allows
this triangular caustic to intersect the source trajectory to produce the
June 19-20 caustic crossings, but only if the orbital velocity exceeds the
escape velocity of the binary system.
  \label{fig-caustic} }

\figcaption[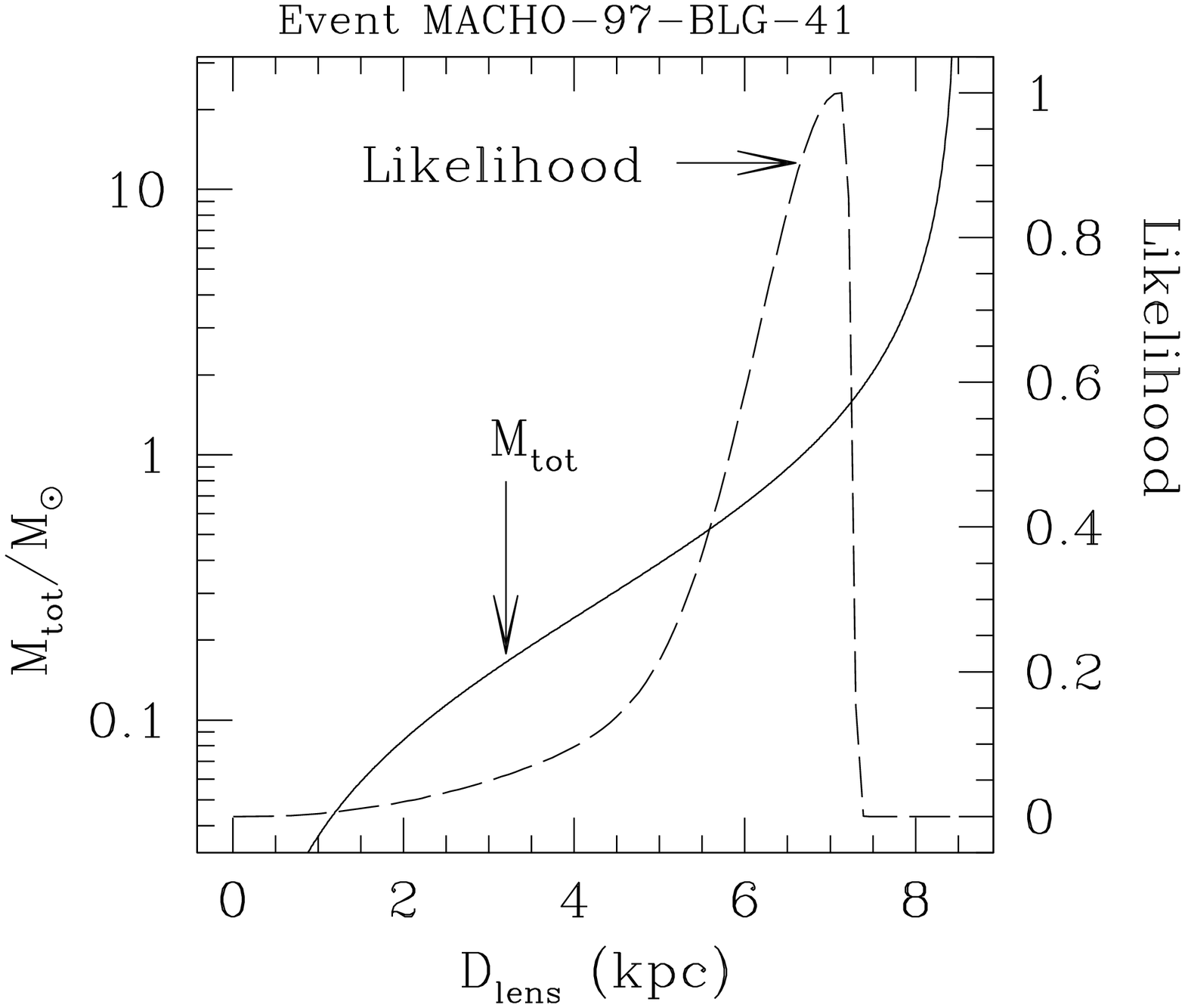]
{The mass-distance relation for the MACHO-97-BLG-41 lens system
is plotted along with a likelihood function for the lens system distance.
The measurements of $\mu$ and $t_E$ provide two constraints
on the three unknowns of a microlensing event
(the lens distance, mass, and transverse velocity), which allows us to 
solve for the single parameter family of solutions shown above.
The transverse velocity of the lens
is also determined as a function of the distance to the lens.
Taking the velocity distributions from
a standard Galactic model\rlap,\cit{han-gould} we are able to construct
a likelihood function for the distance and mass of the lens system.
For a lens distance $\gsim 7\kpc$, the primary lens must be more massive
than the Sun and will contribute a significant amount of unlensed light
superimposed on the source star. Using an upper limit on the brightness
of the primary lens star of $20\pm 10\,$\% of the source brightness
and assuming that the V-band brightness of a main sequence star
$\propto M^{4.5}$ (which is appropriate for $M \gsim 1\,\msun$), we find
the likelihood function shown above. The very
steep cutoff at $D_{\rm lens} > 7\kpc$ is due to this constraint on the
brightness of the lens. Our main sequence assumption could be avoided if the
primary lens was a white dwarf, but since white dwarfs with masses $> 1\,\msun$
are rare, it is unlikely that our limits on the mass and location of
the lens system are violated.
  \label{fig-masslike} }
\pagebreak

\begin{figure}
\plotone{fig_3vs2b.ps}
\end{figure}
\begin{figure}
\plotone{causticfigbw.ps}
\end{figure}
\begin{figure}
\plotone{masslike_97b41ms.ps}
\end{figure}

\end{document}